\documentstyle[prl,aps]{revtex}
\begin{document}
\preprint{DOE/ER/40537-001/NPL94-07-01}
\title{Natural Wormholes as Gravitational Lenses}
\author{
John G.~Cramer$^{(1)}$\thanks{Internet address: cramer@npl.washington.edu},
Robert L.~Forward$^{(2)}$\thanks{Internet address:
forward@sand.npl.washington.edu},
Michael S. Morris$^{(3)}$\thanks{Internet address: msmorris@ovid.butler.edu},\\
Matt Visser$^{(4)}$\thanks{Internet address: visser@kiwi.wustl.edu},
Gregory Benford$^{(5)}$\thanks{Internet address: molsen@vmsa.oac.uci.edu},
and Geoffrey A. Landis$^{(6)}$\thanks{Internet address: glandis@lerc.nasa.gov}}
\address{
$^{(1)}$ Department of Physics FM-15, University of Washington, Seattle WA
98195\\
$^{(2)}$ Forward Unlimited, P.\ O.\ Box 2783, Malibu CA 90265\\
$^{(3)}$ Department of Physics and Astronomy, Butler University, Indianapolis
IN 46208\\
$^{(4)}$ Physics Department, Washington University, St.\ Louis MO 63130-4899\\
$^{(5)}$ Physics Department, University of California at Irvine, Irvine CA
92717-4575\\
$^{(6)}$ NASA Lewis Research Center, Mail Code 302-1, Cleveland OH 44135-3191}
\date{28 June 1994}
\maketitle
\begin{abstract}
Visser has suggested traversable 3-dimensional wormholes that
could plausibly form naturally during Big Bang inflation.  A wormhole
mouth embedded in high mass density might accrete mass, giving the
other mouth a net {\em negative} mass of unusual gravitational
properties.  The lensing of such a gravitationally negative anomalous
compact halo object (GNACHO) will enhance background stars with a
time profile that is observable and qualitatively different from
that recently observed for massive compact halo objects (MACHOs)
of positive mass.  We recommend that MACHO search data be analyzed
for GNACHOs.
\end{abstract}

\section{Introduction: Wormholes and Negative Mass}\label{sec:intro}

The work of Morris and Thorne\cite{Mo:88,Mo:88a} has led to a great
deal of interest in the formation and properties of three-dimensional
wormholes (topological connections between separated regions of
space-time) that are solutions of the Einstein's equations of
general relativity.  Subsequently Visser\cite{Vi:89} suggested a
wormhole configuration, a flat-space wormhole that is framed by
``struts'' of an exotic material, a variant of the cosmic string
solutions of Einstein's equations\cite{Vi:84,Go:85}.  To satisfy
the Einstein field equations the cosmic string framing Visser
wormholes must have a {\em negative} string tension\cite{Vi:89} of
$-1/4G$ and therefore a negative mass density.  However, for the
total mass of the wormhole system, the negative mass density of
the struts should be combined with the effective positive mass
density of the wormhole's gravitational field.  The overall object
could, depending on the details of the model, have positive, zero,
or negative net external mass.  Note that in hypothesizing the
existence of such a wormhole, one has to abandon the averaged null
energy condition\cite{Mo:88,Mo:88a}.  Therefore, the hypotheses
underlying the positive mass theorem no longer apply and there is
nothing, in principle, to prevent the occurrence of negative total
mass\cite{Vi:89a}.  Some of the Visser wormhole configurations have
the shape of a cube or other geometrical solid, but one particularly
simple configuration is a flat-space wormhole mouth framed by a
single continuous loop of exotic cosmic string.

It has been suggested\cite{Vi:84,Go:85} that the inflationary phase
of the early universe might produce closed loops of cosmic string.
It is therefore at least plausible that a similar mechanism might
produce negative-mass string loops framing stable Visser wormholes.

If a particle with positive electric charge passes through such a
wormhole, its lines of force, threading through the wormhole
aperture, give the entrance mouth an effective positive charge
(flux lines radiating outward) and give the exit mouth an effective
negative charge (flux lines converging inward.) Similarly, when a
massive object passes through the wormhole, the same back-reaction
mechanism might cause the entrance mouth to gain mass and the exit
mouth to lose mass\cite{Fr:90}.  Now let us consider a stable Visser
wormhole with  near-zero mass residing in the mass-energy rich
environment of the early universe.  The expected density fluctuations
of the early universe suggest that the separated wormhole mouths
will reside in regions of differing mass density, leading to a mass
flow between the regions they connect.  As this mass passes through
the wormhole, the entrance mouth will gain mass while the exit
mouth will lose mass by the same amount.  Soon, if the mass flow
continues, the exit wormhole mouth will acquire a net negative
mass.

This will lead to a gravitational instability, since the positive
mass mouth will attract more mass through its aperture while the
negative mass mouth will gravitationally repel nearby mass\cite{Bo:57}.
Thus the positive-negative imbalance will be fed by gravity and
will continue to grow.  If this process proceeds without interruption,
the exit wormhole mouth might develop a stellar-scale negative
mass.  Visser wormholes, therefore, provide at least one motivation
for seriously considering the possible existence of naturally
occurring astronomical objects of sizable negative mass.

We have tried to sketch only one scenario for the possible existence
of negative mass objects.  We must, however, point out that the
non-existence of such objects would not rule out the existence of
natural wormholes, since their properties are highly model-dependent.
Also, observations providing evidence of negative mass objects
would not require the existence of natural wormholes, since other
negative-mass objects could conceivably exist, but would indicate
a direct violation of the averaged null energy condition mentioned
above.

Negative-mass objects, while repelling all nearby mass (positive
or negative), would themselves be attracted\cite{Bo:57} by the mass
of a nearby galaxy and might form part of a galactic halo.  They
would have unusual gravitational properties that could produce
detectable gravitational lensing effects.  These lensing effects,
for the same absolute mass, are of the same magnitude as those
recently detected for massive cosmic halo objects
(MACHOs)\cite{Al:93,Au:93}, but, as we will show below, are
qualitatively different in shape.  We here examine in detail the
lensing effects of gravitationally negative anomalous compact halo
objects (GNACHOs).

\section{Gravitational Lensing by Negative Masses}\label{sec:lens}

Naively, if a gravitationally attractive positive mass acts as a
converging lens that brightens a background star, one might expect
a gravitationally repulsive negative mass to act as a diverging
lens that causes a background star to briefly grow dimmer.  Actually
the lensing of a negative mass is not analogous to a diverging
lens.  In certain circumstances it can produce more light enhancement
than does the lensing of an equivalent positive mass.

Fig.\ \ref{fig:1} shows the geometry of starlight that is
gravitationally lensed by an object of negative mass.  Starlight
is radiated from the stellar source $S$ and is detected by the
observer at $D$.  A gravitationally negative object $N$ lies between
source and observer at an impact-parameter distance $b$ from the
source-detector axis $DS$.   An off-axis ray of light passes near
the negative mass object $N$, coming within a distance $a$ of it,
and is deflected\cite{Go:74,Pa:86} by an angle $\delta$ = $(4 G
|M| / c^{2}) / a$ where $G$ is Newton's gravitational constant,
$|M|$ is the absolute value of the deflecting mass, and $c$ is the
velocity of light.  The angle $\delta$ is an external angle of the
triangle formed by the direct and deflected rays.  This triangle
has interior angles $\alpha$ and $\beta$, so $\delta$ = $\alpha +
\beta$.   If the detector-source distance is $L_{S}$ and the
detector-mass distance is $L_{N}$, then we have the following
equations:
\begin{eqnarray}\label{eq:1}
\alpha = (b - a)/L_{N}~~~~(\alpha \ll 1)\\
\beta  = (b - a)/(L_{S} - L_{N})~~~~(\beta \ll 1)\\
\delta = \alpha + \beta = \frac{4 G |M_{N}|}{c^{2}}\,\frac{1}{a}.
\end{eqnarray}
These lead to the dimensionless quadratic equation:
\begin{equation}\label{eq:4}
A^2 - A B + 1 = 0~~~~~(B = A + 1/A)
\end{equation}
where $A \equiv a / a_{0}$ is the dimensionless distance of closest
approach of the deflected ray, $B \equiv b / a_{0}$ is the
dimensionless impact parameter distance of the deflecting mass,
and
\begin{equation}\label{eq:5}
a_{0} \equiv \sqrt{\frac{4 G |M_{N}|}{c^{2}}\,
                   \frac{L_{N} (L_{S} - L_{N})}{L_{S}}}
\end{equation}
is the characteristic gravitational length scale of the problem.
For a positive lensing mass, $a_{0}$ would be the radius of the
Einstein ring produced when the mass is positioned at zero impact
parameter ($b$=0).  To give some feeling for this length scale, if
the stellar source $S$ is in the Large Magellanic Cloud and a
negative lensing mass $N$ of one solar mass is in the galactic
halo, then $L_{S}$ = 2 $\times$ 10$^{21}$~m, $L_{N}$ = 5 $\times$
10$^{20}$~m, and $a_{0}$ = 1.5 $\times$ 10$^{12}$~m or about 10~AU.

Solving quadratic (\ref{eq:4}) for $A$ gives two solutions:
\begin{equation}\label{eq:6}
A_{\pm} = \frac{1}{2} [B \pm \sqrt{B^{2} - 4}].
\end{equation}
When $B > 2$ there are two real solutions of the quadratic,
corresponding to two rays that are deflected to the observer.  When
$B < 2$ there are no real solutions, indicating that the deflection
is blocking all rays from reaching the observer.  At $B$ = 2,
$A_{+}$ = $A_{-}$ = $B/2$ and, as will be discussed below, the
rainbow-like caustic occurs, allowing many rays to reach the observer
and producing a dramatic brightening of the background star.  Fig.
\ref{fig:2} shows this schematically. The negative mass deflects
rays in inverse proportion to their distance of closest approach,
creating a shadowed umbra region where light from the source is
extinguished.  At the edges of the umbra the light rays accumulate
to form the caustic and give a very large increase in light intensity.
The intensity falls slowly to normal at larger transverse distances.

\section{Light Modulation Profiles}\label{sec:profile}

If the unmodified intensity of the background star is $I_{0}$ and
the altered intensity of the background star in the presence of
the negative lensing mass for each solution is $I_{\pm}$, then the
partial amplification factors $p_{\pm} \equiv I_{\pm}/I_{0}$ are
given by\cite{Go:74}:
\begin{equation}\label{eq:7}
p_{\pm} = \frac{a_{\pm}}{b}\,\left|\frac{d a_{\pm}}{d b}\right|
= \frac{A_{\pm}}{B}\,\left|\frac{d A_{\pm}}{d B}\right|
= \frac{(B \pm \sqrt{B^{2} - 4})^{2}}{4 B \sqrt{B^{2} - 4}}\\
\end{equation}
The overall relative intensity ${\cal I}_{neg} = p_{+} + p_{-}$ is
the modulation in brightness of the background star as detected by
the observer, and is given by:
\begin{equation}\label{eq:8}
{\cal I}_{neg} = p_{+} + p_{-} = \frac{B^{2} - 2}{B \sqrt{B^{2} - 4}}.\\
\end{equation}

It is interesting to compare this with the similar expression for
relative intensity ${\cal I}_{pos}$ of a background source that is
lensed by an object of {\em positive} mass:
\begin{equation}\label{eq:9}
{\cal I}_{pos} = \frac{B^{2} + 2}{B \sqrt{B^{2} + 4}}.
\end{equation}
For the same dimensionless impact parameter $B$ with $B > 2$, it
is always true that ${\cal I}_{neg} > {\cal I}_{pos}$, so for large
impact parameters a negative mass actually provides {\em more}
light enhancement through lensing than an equivalent positive mass.
When $B \rightarrow 2$, the overall intensity ${\cal I}_{neg}
\rightarrow \infty$, a condition indicating a caustic at which
light rays from many trajectories are deflected toward the observer.
It provides a distinctive and unusual signature for negative-mass
lensing.

The intensity modulation that is actually observed occurs when the
lensing mass, which is assumed to be moving with transverse velocity
$V$, crosses near the source-detector axis $DS$ with a minimum
impact parameter $b_{0}$ and a minimum dimensionless impact parameter
$B_{0}$ = $b_{0}/a_{0}$.  The time-dependent impact parameter is
therefore $b(t) = \sqrt{b_{0}^2 + (V t)^2}$, and
\begin{equation}\label{eq:10}
B(t) = B_{0} \sqrt{1 + (\frac{t}{T_{0}})^{2}}
\end{equation}
where $T_{0}$ = $b_{0}/V$ is the transit time across the distance
of the minimum impact parameter and is the characteristic time
scale of the problem.

Eqns.\ \ref{eq:8} and \ref{eq:10} are used to calculate the light
enhancement profile of the process, taking ${\cal I}_{neg}$ = 0
when $|B| < 2$, and there are no real solutions to Eqn.\ \ref{eq:4}.
Fig.\ \ref{fig:3} shows these light enhancement profiles, plotting
${\cal I}_{neg}$ vs. $t/T_{0}$ for a range of minimum dimensionless
impact parameters ranging from $B_{0}$ = 0.50 to 2.20.  Fig.\ \ref{fig:4}
shows similar light enhancement profiles for a positive mass of
the same size and range of $B_{0}$ values.  As can be seen, the
light enhancement profiles are qualitatively different for positive
and negative lensing masses of the same magnitude and geometry.
In particular, the negative mass curves are much sharper, show
stronger but briefer light enhancements, and for $B_{0} < 2$ show
a precipitous drop to zero intensity, i. e., extinction of the
light from the source when the lensing mass deflects all light from
$S$ away from the observer.

Such a signature might possibly be confused with that of occultation
by a dark foreground object.  However, an occultation would not be
preceded by a dramatic rise in intensity and might, if the object
had a significant atmosphere, have different light profiles at
different wavelengths.  Therefore, the negative gravitational
lensing presented here, if observed, would provide distinctive and
unambiguous evidence for the existence of a foreground object of
negative mass.

\section{Conclusion}\label{sec:conclusion}

The calculations presented above show that objects of negative
gravitational mass, if they exist, can provide a very distinctive
light enhancement profile.  Since three groups are presently
conducting searches for the gravitational lensing of more normal
positive mass objects, we suggest that these searches be slightly
broadened so that the signatures of the objects discussed above
are not overlooked by over-specific data selection criteria and
software cuts.  We recommend that MACHO search data be analyzed
for evidence of GNACHOs.

\acknowledgements

This work was initiated at the NASA Jet Propulsion Laboratory (JPL)
Workshop on Advanced Quantum/Relativity Theory Propulsion Concepts
organized by Dr. Robert H. Frisbee of JPL and sponsored by Dr. Gary
L. Bennett of the NASA Office of Advanced Concepts and Technology.
The authors wish to thank the other workshop participants for
stimulating discussions and useful suggestions.  The work of JGC
was supported in part by the Division of Nuclear Sciences of the
U.~S.~Department of Energy under Grant DE-FG06-90ER40537.  The work
of RLF was supported in part by NASA through JPL.  The work of MV
was supported in part by the U.~S.~Department of Energy.

\begin{figure}
\caption{Geometry for gravitational lensing by a negative mass
object.  Off axis light rays from stellar source $S$ are deflected
to detector $D$ by the gravitational repulsion of the negative
lensing mass (GNACHO) $N$.}
\label{fig:1}
\end{figure}

\begin{figure}
\caption{Light deflection by a negative mass object (horizontal
scale highly compressed).  Light is swept out of the central region,
creating an umbra region of zero intensity.  At the edges of the
umbra the rays accumulate, creating a rainbow-like caustic and
enhanced light intensity.}
\label{fig:2}
\end{figure}

\begin{figure}
\caption{Intensity profile of a gravitationally negative anomalous
compact halo object (GNACHO) as it passes near the source-detector
axis $DS$.  The several curves correspond to minimum dimensionless
impact parameter values $B_{0}$ = 0.50 (at edge of plot), 0.75,
1.00, 1.25, 1.50, 1.75, 2.00, 2.10, and 2.20 (small central bump).
(See text for definitions of the variables.)}
\label{fig:3}
\end{figure}

\begin{figure}
\caption{Intensity profile of an object of equivalent positive mass
in the same geometries.  Here $B_{0}$ = 0.50 is the highest curve,
and $B_{0}$ = 2.20 is the lowest.}
\label{fig:4}
\end{figure}

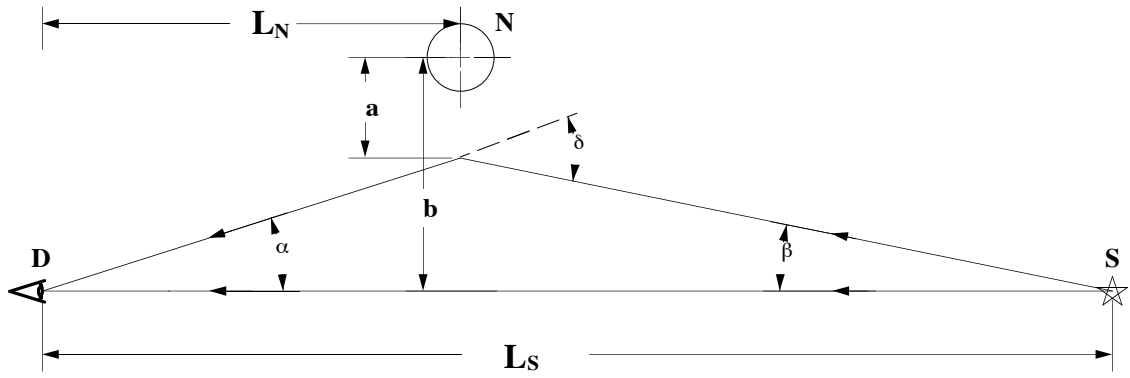
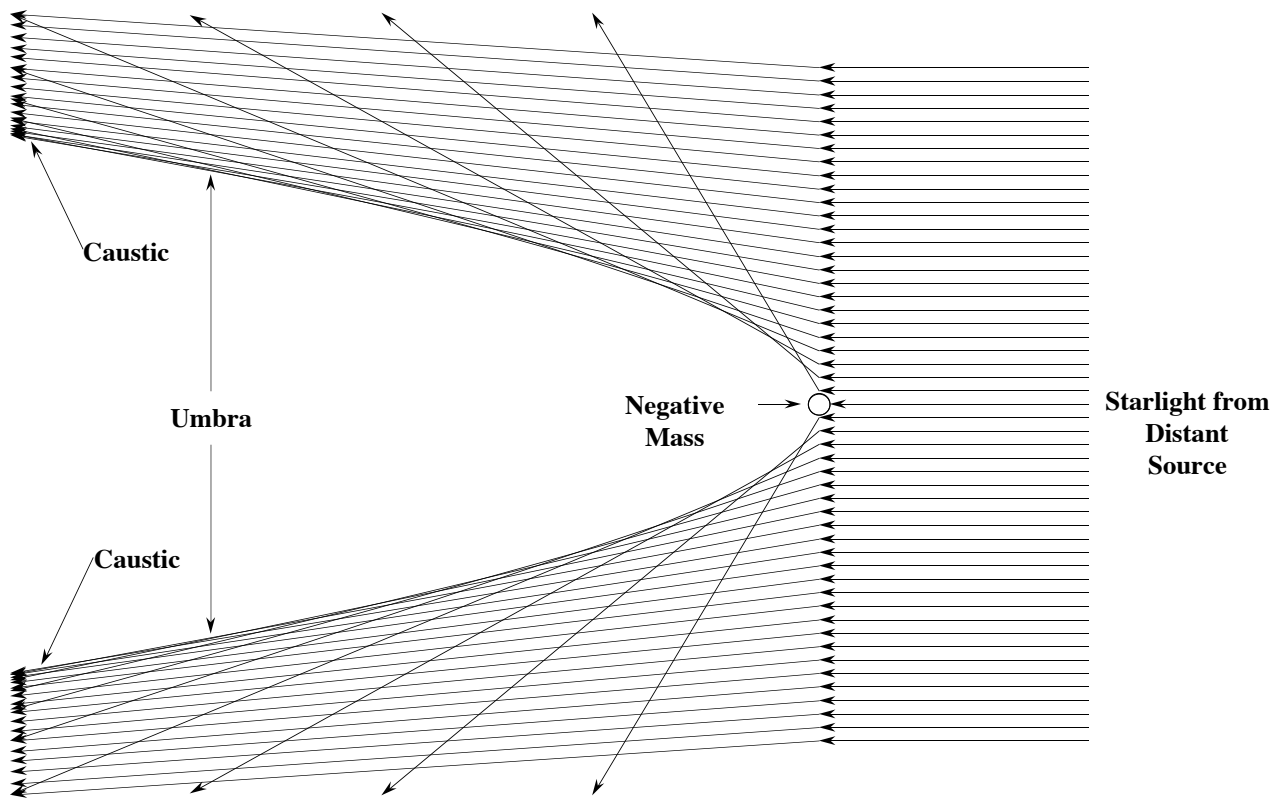
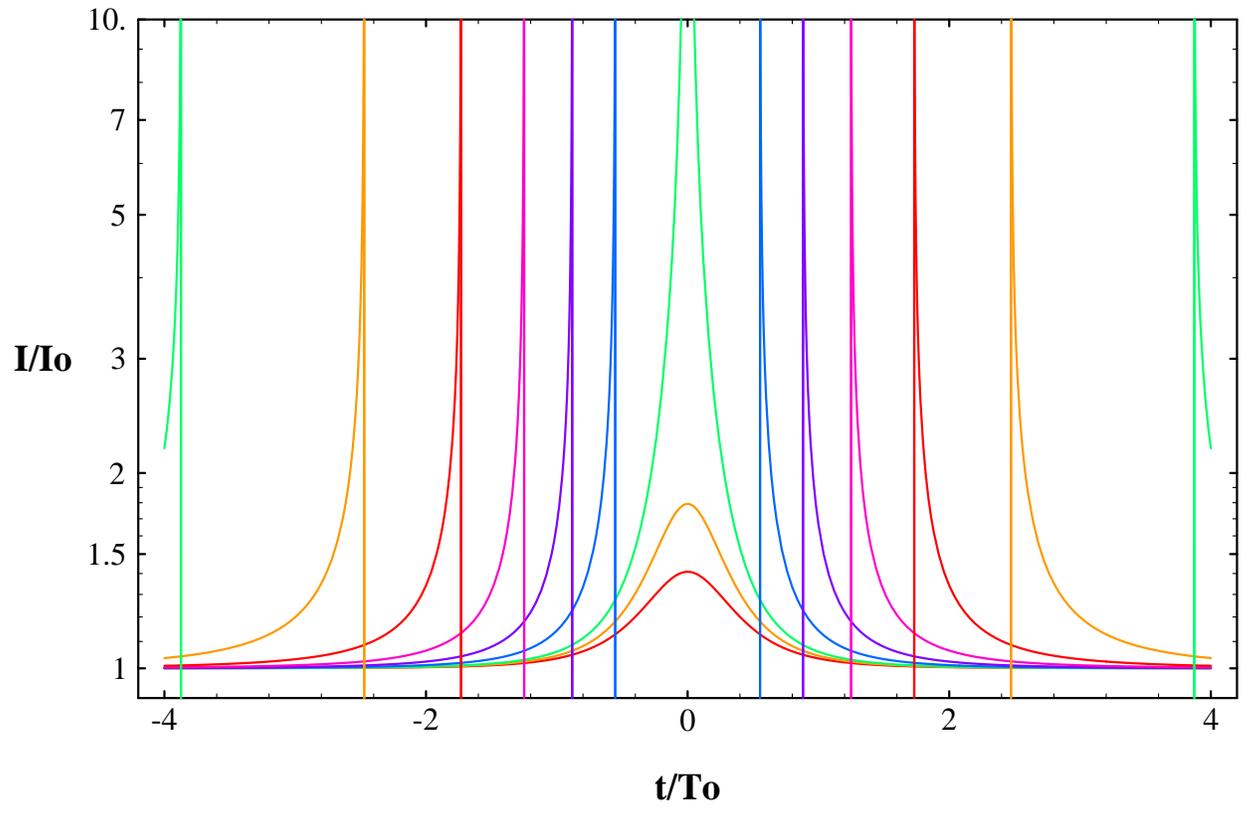
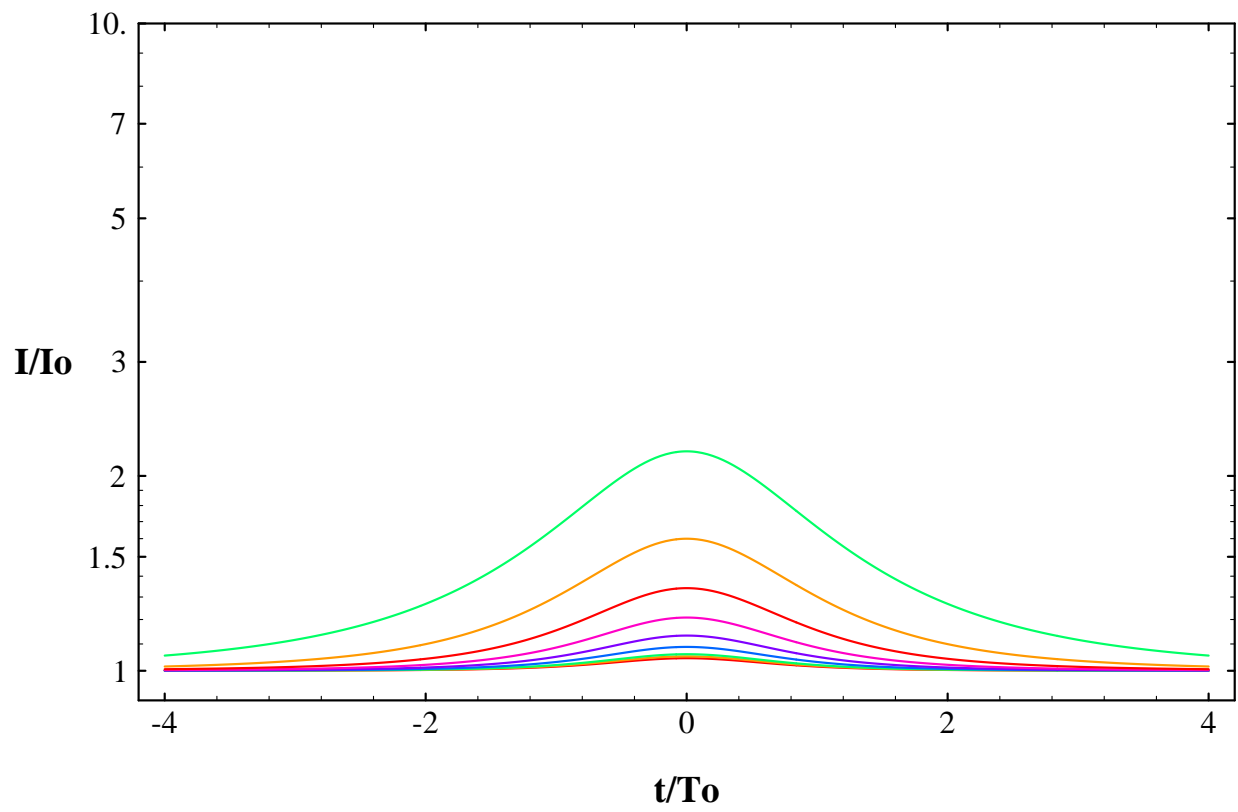
\end{document}